\documentclass[aps,prl,twocolumn,showpacs,superscriptaddress,amsmath,amssymb]{revtex4}
\usepackage{amsfonts}
\usepackage{amsmath}
\usepackage{amssymb}
\usepackage{graphicx}%
\usepackage{color}
\providecommand{\U}[1]{\protect\rule{.1in}{.1in}}
\newtheorem{theorem}{Theorem}

\newtheorem{lemma}[theorem]{Lemma}

\definecolor{nblue}{rgb}{0.2,0.2,0.7}
\definecolor{ngreen}{rgb}{0.2,0.6,0.2}
\definecolor{nred}{rgb}{0.7,0.2,0.2}
\definecolor{nblack}{rgb}{0,0,0}

\begin{document}
%
%
%
%
%
%

\title{}
\title{  The  asymmetry properties of pure quantum states }
\author{Iman Marvian}
\affiliation{Perimeter Institute for Theoretical Physics,  Waterloo,
Ontario, Canada N2L 2Y5}

\affiliation{Institute for Quantum Computing, University of Waterloo, Waterloo, Ontario, Canada N2L 3G1}
\affiliation{Department of Physics and Astronomy, Center for Quantum Information Science and Technology, University of Southern California, Los Angeles, CA 90089}

\author{Robert W. Spekkens}
\affiliation{Perimeter Institute for Theoretical Physics,  Waterloo,
Ontario, Canada N2L 2Y5}

\date{\today}

\begin{abstract}
The \emph{asymmetry properties} of a state relative to some symmetry group specify how and to what extent the given symmetry is broken by the state.   Characterizing these is found to be surprisingly useful for addressing a very common problem: to determine what follows from a system's dynamics (possibly open) having that symmetry.  We demonstrate and exploit the fact that the asymmetry properties of a state can be understood in terms of information-theoretic concepts. We show that for a pure state $\psi$ and a symmetry group $G$, they are completely specified by the characteristic function of the state, defined as $\chi_{\psi}(g)\equiv \langle \psi|U(g)|\psi\rangle$ where $g\in G$ and $U$ is the unitary representation of interest. Based on this observation, we study several important  problems about the interconversion of pure states under symmetric dynamics such as determining the conditions for reversible transformations, deterministic irreversible transformations and asymptotic transformations.

\end{abstract}

\maketitle

\setcounter{section}{0}

\setcounter{equation}{0}
Suppose that the only thing one knows about a complicated quantum dynamics, which is possibly open, is that it has a particular symmetry.  What does this imply about the evolution of the system's state?  Alternatively, suppose one is given a description of an initial  quantum state and a possible final state for a system. Can the first evolve to the second by symmetric dynamics?  These sorts of problems arise in many physical contexts.  For instance, they are clearly important in any situation wherein one might apply Noether's theorem (which infers conservation laws from symmetries in the case of closed dynamics).  To answer them, it is useful to study the \emph{asymmetry properties of a state}, that is, those properties which specify how and to what extent the given symmetry is broken by the state.  If the dynamical equations are  invariant under a symmetry group of transformations then there are constraints on how the asymmetry properties can change. For instance, the  final state can only break the symmetry in ways in which it was broken by the initial state, and its measure of asymmetry can be no greater than that of the initial state \cite{MS11,Pure}. In other words, \emph{symmetric dynamics cannot generate asymmetry}.

Furthermore, if one is in a scenario wherein implementing symmetric dynamics is easy while implementing dynamics that break the symmetry is hard or impossible then asymmetric states become a resource (for instance in the case where two parties lack a shared reference frame \cite{BRS07}.).  
 Indeed, developing the theory of how the resource of asymmetry can be quantified and manipulated provides another useful angle on the problem of determining the consequences of symmetric dynamics. Such a resource theory is analogous to entanglement theory: the constraint of symmetric dynamics is the analogue of the constraint of local operations and classical communication and asymmetric states are the analogues of entangled states.  For almost any question that one might pose about entanglement, one can ponder the analogous question for asymmetry.  The resource perspective has been an extremely  useful method for organizing results about entanglement, so one may expect the same to be true of asymmetry as well.

In this article, we answer the most basic of such resource theory questions: what are the conditions under which two  quantum states can be converted one to the other reversibly  under symmetric operations?  What are the conditions if the conversion is not required to be reversible?  What are the conditions under which many copies of one state can be (approximately) reversibly interconverted to many copies of another and what is the rate of interconversion? We consider only interconversion of pure quantum states in this article. Such questions have been considered previously, for instance in Refs.~\cite{GS07,GMS09,TGS11}.
However, whereas previous work attacked the problem for  one or two  particular symmetry groups, most of our results apply to arbitrary compact Lie groups and finite groups. Our results  are  therefore of much greater generality and this generality clarifies how they ought to be interpreted.  


Another important motivation comes from the field of \emph{quantum metrology}, wherein one explores the use of quantum techniques to achieve greater precision for a variety of different kinds of parameter estimation tasks \cite{Metrology}. High-precision clocks, gyroscopes and accelerometers are prominent examples, for which achieving a quantum improvement in precision would have significant applications for the rest of physics. The parameter to be estimated for such tasks is an unknown element of a group.  For instance, the task of aligning a pair of Cartesian reference frames by transmitting a system that breaks rotational symmetry and estimating its orientation is clearly of this sort (see \cite{BRS07} for a review of this topic).  The degree of success one can achieve in any such task is clearly a function of the asymmetry properties of the state that is transmitted, so a systematic study of these properties can help to develop optimal protocols and strategies for dealing with practical constraints such as noise.

For more discussions of the ideas presented here, and for both proofs and lengthier expositions of the results, see Refs.~\cite{MS11,Pure, MS11part2}.

We describe  time evolutions  by quantum channels, i.e.  linear, trace preserving, completely positive super-operators which map density operators in the input Hilbert space, $\mathcal{H}_{in}$, to density operators in the output Hilbert spaces, $\mathcal{H}_{out}$, where in general those can be different spaces. This way of describing a time evolution is general enough to include closed system dynamics as well as all the open system dynamics in which the environment is initially uncorrelated with the system.

Now consider a  symmetry which is described by a group $G$ and its unitary representations on  $\mathcal{H}_{in}$ and $\mathcal{H}_{out}$,  $\{U_{\textrm{in}}(g),  g\in G\}$ and $\{U_{\textrm{out}}(g),  g\in G\}$ respectively. Then the time evolution described by quantum channel $\mathcal{E}$ has the symmetry $G$, or is  \textbf{G-covariant}, iff
\begin{equation}
\forall g\in G:\  \mathcal{E}\left(U_{\textrm{in}}(g)(\cdot)U_{\textrm{in}}^\dag(g)\right)=U_{\textrm{out}}(g)\mathcal{E}\left(\cdot\right)U_{\textrm{out}}^\dag(g).
\end{equation}
We here focus on the case of compact Lie groups and finite groups. If there is a G-covariant time evolution under which the state  $\rho$ evolves to the state $\sigma$, we denote it by $\rho\xrightarrow{G-cov} \sigma$.  As we mentioned, $\rho$ and $\sigma$ can be density operators on different Hilbert spaces, but without loss of generality we can always assume these two Hilbert spaces are two different sectors of a larger Hilbert space $\mathcal{H}\equiv\mathcal{H}_{in}\oplus \mathcal{H}_{out}$ where the representation of $G$ is  $\{U(g)\equiv U_{in}(g)\oplus U_{out}(g) :g\in G\}$ (see appendix A of \cite{MS11}).


The G-covariant time evolutions  define equivalence classes of states and the asymmetry properties of a state are precisely those that are necessary and sufficient to determine its equivalence class. We will say that \emph{two states $\rho$ and $\sigma$ have exactly the same \emph{asymmetry properties} (with respect to the group $G$) or they are  \textbf{G-equivalent} if they are reversibly interconvertible by $G$-covariant operations, i.e., $\rho\xrightarrow{G-cov} \sigma$ and $\sigma\xrightarrow{G-cov} \rho$.}  Let  $\rho$ and $\sigma$ be two G-equivalent states; then    $\rho\xrightarrow{G-cov} \tau$ implies  that $\sigma\xrightarrow{G-cov} \tau$. In other words, to determine whether there exists a G-covariant time evolution which transforms one state  to another, the only thing we need to know is the G-equivalence class of these two states.  Note that the  G-equivalence class of a state also specifies all the symmetries of the state, i.e. if $\rho$ and $\sigma$ are G-equivalent then if for some group element $g\in G$, $U(g)\rho U^{\dag}(g)=\rho$ then $U(g)\sigma U^{\dag}(g)=\sigma$.  We denote all the group elements under which $\rho$ is invariant by $\textrm{Sym}_G(\rho)$ \footnote{This is the subgroup of $G$ that stabilizes $\rho$.}.

We also introduce an equivalence relation over states that is slightly stronger than G-equivalence: \emph{Two pure states, $\psi$ and $\phi$, are called \textbf{unitarily G-equivalent}  if there exists a  unitary $V_{\textrm{G-inv}}$  such that
$\forall g\in G: [V_{\textrm{G-inv}},U(g)]=0$ and $V_{\textrm{G-inv}} |\psi\rangle= |\phi \rangle$.} Such a unitary is called a \textbf{G-invariant unitary}.  Note that if two pure states are unitarily G-equivalent, then they are also G-equivalent but
the opposite implication does not hold.

 The above definition of asymmetry properties   is based on the intuition that asymmetry is something which cannot be generated by symmetric time evolutions.  We call this the \emph{constrained-dynamical} perspective. However, one can also take an \emph{information-theoretic} perspective on how to define the asymmetry properties of a state.  

  Consider a set of communication protocols in which one chooses a message $g\in G$ according to a measure over the group and then sends the state $\mathcal{U}(g)[\rho]$ where $\rho$ is some fixed state. The goal of the sender is to inform the receiver about the specific chosen group element. We claim that the asymmetry properties of a state $\rho$ can be defined as those that determine  the effectiveness of using the signal states $\{ \mathcal{U}(g)[\rho] : g\in G \}$ to communicate a message $g\in G$. To get an intuition for this, note that if $\rho$ is invariant under the effect of some specific group element $h$ then the state used for encoding $h$ would be the same as the state used for encoding the identity element $e$, ($\mathcal{U}(h)[\rho]=\mathcal{U}(e)[\rho]=\rho$), such that the message $h$ cannot be distinguished from $e$.
  In the extreme case where $\rho$ is invariant under all group elements this encoding does not transfer any information. 

 So from this point of view, the asymmetry properties of $\rho$ can be inferred from the information-theoretic properties of the encoding $\{\mathcal{U}(g)[\rho]: g\in G\}$.  To compare the asymmetry properties of two arbitrary states  $\rho$ and $\sigma$, we have to compare the information content of two different encodings: $\{\mathcal{U}(g)[\rho]: g\in G\}$ (encoding I) and $\{\mathcal{U}(g)[\sigma]: g\in G\}$ (encoding II).   If each state $\mathcal{U}(g)[\rho]$ can be converted to   $\mathcal{U}(g)[\sigma]$ for all $g\in G$, then encoding I has as much or more information about $g$ than encoding II.  If the opposite conversion can also be made, then the two encodings have precisely the same information about $g$.  Consequently, in an information-theoretic characterization of the asymmetry properties, it is the reversible interconvertability of the sets (defined by the two states) that defines equivalence of their asymmetry properties.

As it turns out, our two different approaches lead to the same definition of asymmetry properties, as the following lemmas imply.
\begin{lemma}
The following statements are equivalent:\\
\textbf{A})\ There exists a G-covariant quantum channel $\mathcal{E}_{\textrm{G-cov}}$ 
such that $\mathcal{E}_{\textrm{G-cov}}(\rho)=\sigma$\\
\textbf{B})\  There exists a  quantum channel $\mathcal{E}$ 
such that $\forall g\in G:\ \ \   \mathcal{E}(\mathcal{U}(g)[\rho])=\mathcal{U}(g)[\sigma].$
\end{lemma}
\begin{lemma}\label{unitary}
The following statements are equivalent:\\
\textbf{A})\ There exists a G-invariant unitary  $V_{\textrm{\textrm{G-inv}}}$ such that $V_{\textrm{G-inv}}|{\psi}\rangle=|{\phi}\rangle$ \\
\textbf{B})\ There exists a unitary $V$ such that
$\forall g\in G:\ \ \   V U(g)|{\psi}\rangle=U(g)|{\phi}\rangle.$
\end{lemma}

 Note that in both of these lemmas, the condition $\textbf{A}$ concerns whether it is possible to transform a single state to another under a limited type of dynamics.   On the other hand, in the \textbf{B} condition, there is no restriction on the dynamics, but now we are asking whether one can transform a \emph{set} of states to another set.

In the following, we find the characterization of the unitary G-equivalence classes  for pure states via both of these points of view, i.e. constrained-dynamical and information-theoretic. We start with the constrained-dynamical point of view. Suppose $\{U(g):g\in G \}$ is a representation of a finite or compact Lie group $G$ on the Hilbert space $\mathcal{H}$.  We can always decompose this representation to a discrete set of finite dimensional irreducible representations (irreps).  This suggests the following decomposition of the Hilbert space,
$\mathcal{H}=\bigoplus_{\mu} \mathcal{M}_{\mu}\otimes  \mathcal{N}_{\mu}$
where $\mu$ labels the irreps and $\mathcal{N}_{\mu}$ is the subsystem associated to the multiplicities of representation $\mu$ (the dimension of $ \mathcal{N}_{\mu}$ is equal to the number of multiplicities of the irrep $\mu$ in this representation). Then the effect of $U(g)$ can be written as $U(g) =\bigoplus_\mu U_\mu(g) \otimes I_{\mathcal{N}_\mu}$ where $U_\mu(g)$ acts on $\mathcal{M}_{\mu}$ irreducibly and where $I_{\mathcal{N}_\mu}$ is the identity operator on $\mathcal{N}_{\mu}$.  Using this decomposition  and Schur's lemmas, one can show that any arbitrary G-invariant unitary is of the following form \cite{BRS07},
$V_{\textrm{G-inv}}=\bigoplus_\mu I_{\mathcal{M}_\mu} \otimes V_{\mathcal{N}_\mu},$
where  $V_{\mathcal{N}_\mu} $ acts unitarily on ${\mathcal{N}_\mu} $. We can then easily prove the following theorem (Here  $\Pi_{\mu}$ is the projection operator onto the subspace  $\mathcal{M}_{\mu}\otimes  \mathcal{N}_{\mu}$, the subspace associated to the irrep $\mu$.):
\emph{Two pure states $\psi$ and $\phi$ are unitarily G-equivalent iff}
\begin{equation} \label{reduction}
\forall \mu:\  \mathrm{tr}%
_{\mathcal{N}_{\mu}}({\Pi}_{\mu}\ |\psi\rangle\langle\psi|\ {\Pi}_{\mu})=\mathrm{tr}_{\mathcal{N}_{\mu}}({\Pi}_{\mu}\ |\phi\rangle\langle\phi|{\Pi
}_{\mu})
\end{equation}
For an arbitrary pure state $\psi$, we call the set of operators $\{\rho^{(\mu)}\equiv \text{tr}_{ \mathcal{N}_{\mu}} ({\Pi}_{\mu}\  |\psi\rangle\langle\psi| \ {\Pi}_{\mu} )\}$ the \emph{reduction onto irreps} of $\psi$. So in the above theorem we have proven that the unitary G-equivalence class of a pure state is totally specified by its reduction onto irreps.  
Also, as  is shown in \cite{MS11}, we can generalize the result by showing that for any pair of pure states $\psi_{1},\psi_{2}$  if the  distance between the reductions is small, then there exists a G-invariant unitary $V$ such that the fidelity  $|\langle\psi_2|V|\psi_1\rangle|$ is large, and if  the  distance between the reductions is large, then for all unitaries $V$,  $|\langle\psi_2|V|\psi_1\rangle|$ is small.


Now we switch to finding the characterization of the unitary G-equivalence classes using the information-theoretic point of view.  Lemma \ref{unitary} implies that $\psi$ and $\phi$
are unitarily G-equivalent iff there is a unitary $V$ such that $\forall g\in G: VU(g)|\psi\rangle=U(g)|\phi\rangle$. Now recall that  there exists a unitary operator $W$ which transforms  $\{{\psi_i}\}$ to $\{{\phi_i}\}$, that is, $\forall i:\  W|{\psi_i}\rangle=|{\phi_i}\rangle$, iff the Gram matrices of the two sets of states are equal, i.e., $\langle\psi_{i}|\psi_{j}\rangle=\langle\phi_{i}|\phi_{j}\rangle$ for all $i,j$ (see e.g.  Ref.~\cite{Jozsa-Schlienz}).  
 So the necessary and sufficient condition for the existence of a unitary $V$ such that $\forall g\in G: VU(g)|\psi\rangle=U(g)|\phi\rangle$ is the equality of the Gram matrices of the set  $\{U(g)|{\psi}\rangle:g\in G\}$  and the set $\{U(g)|{\phi}\rangle :g\in G\}$.  Given that these are, respectively,
$\langle\psi|U^\dag(g_1)U(g_2)|{\psi}\rangle= \langle\psi| U(g^{-1}_1g_2)|{\psi}\rangle$
and
$\langle\phi|U^\dag(g_1)U(g_2)|{\phi}\rangle= \langle\phi|U(g^{-1}_1g_2)|{\phi}\rangle,$
their equality is equivalent to
\begin{equation} \label{Unit-Equiv}
\forall g\in \text{G}:\ \ \langle\psi|U(g)|\psi\rangle=\langle\phi|U(g)|\phi\rangle.
\end{equation}
Motivated by this, we define \emph{the \textbf{characteristic function} of a pure state $\psi$ relative to a unitary representation $\{U(g): g\in G\}$ of a group G as a function $\chi_\psi : G \to \mathbb{C}$ of the form}
$\chi_\psi(g)\equiv\langle\psi|U(g)|{\psi}\rangle.$
To summarize, we have proven that:
\emph{Two pure states $\psi$ and $\phi$ are unitarily
G-equivalent iff $\forall g\in G: \chi_{\psi}(g)=\chi_{\phi}(g)$.}

So we have found two different characterizations of the unitary G-equivalence classes: the reduction onto irreps and the characteristic function. But how are these  related?  It turns out that the connection is via the generalized Fourier transform  over the group. In particular if $\chi_{\psi}$ is the characteristic function of $\psi$ and $\{\rho^{(\mu)}\}$ is its reduction onto irreps, then we have
$\chi_\psi(g)=\sum_\mu \text{tr}(\rho^{(\mu)}U^{(\mu)}(g))$ and $\rho^{(\mu)}=d_\mu \int dg \chi_\psi(g^{-1}) U^{(\mu)}(g)$ where $d_{\mu}$ is the dimension of irrep $\mu$ and $dg$ is the uniform (Haar) measure on the group. (For finite groups we replace the integral with summation.)

Characteristic functions have some nice mathematical properties which make them the preferred way for specifying the unitary G-equivalence classes. In particular we can easily check that:
  1) Characteristic functions multiply under tensor product, \label{multi} ($\chi_{\psi\otimes\phi}(g)= \chi_{\psi}(g)\chi_{\phi}(g)).$
 2) $|\chi_\psi(g)|\leq 1$ for all $g\in G$ and $\chi_\psi(e)=1$ where $e$ is the identity of group.
  3) $|\chi_\psi(g_s)|=1$ for $g_s\in \textrm{Sym}_G(\psi)$.
  4) $|\chi_\psi(g)|=1$ for all $g\in G$ iff $|\psi\rangle\langle\psi|$ is G-invariant i.e. $U(g)|\psi\rangle=e^{i\theta(g)}|\psi\rangle$; in this case  $\chi_{\psi}(g)=e^{i\theta(g)}$ is a 1-d representation of group.

We are now in a position to characterize the G-equivalence classes of states.  Using the above properties of the characteristic function and  the Stinespring dilation theorem for G-covariant channels \cite{KW99}, one can prove:
\emph{For $G$ a compact Lie group, two pure states $\psi$ and $\phi$
are $G$-equivalent iff there exists a 1-dimensional
representation of $G$, $e^{i\Theta(g)}$, such that}
\begin{equation} \label{con-G-Equiv}
\forall g\in G: \chi_{\psi}(g)=\chi_{\phi}(g) e^{i\Theta(g)}
\end{equation}

Comparing to the condition for unitary G-equivalence classes, Eq.(\ref{Unit-Equiv}), here we have an extra phase freedom ($e^{i\Theta(g)}$) for G-equivalence. This extra phase freedom is rooted in the fact that, unlike the case of unitary G-equivalence,  here  the time evolution is not restricted to be unitary and we can couple the system to an ancillary system which is initially in a G-invariant state (and so its characteristic function is a 1-dimensional representation of the group).   Furthermore, based on the fact that the transformation should be reversible, one can argue that the state of the ancillary system after the time evolution  $\psi\xrightarrow{G-cov}\phi$ should still be G-invariant; otherwise one can  build a (Carnot-type) cycle formed by $\psi\xrightarrow{G-cov} \phi$ and $\phi\xrightarrow{G-cov} \psi$ which generates an infinite number of asymmetric states without using any resource. Therefore  if the transformation from $\psi$ to $\phi$ is reversible, then the freedom we get  by using an ancillary system and open G-covariant dynamics is exactly described by  a 1-dimensional representation of the group as in Eq.(\ref{con-G-Equiv}).

The above theorem applies only to the compact Lie groups. In the case of finite groups,  we can  prove Eq.(\ref{con-G-Equiv}) also describes the necessary and sufficient condition for G-equivalence if we make  the extra assumption that the characteristic functions of $\psi$ and $\phi$ are everywhere nonzero.

We have found the condition under which $\psi\xrightarrow{G-cov} \phi$ and $\phi\xrightarrow{G-cov}\psi$. The case of non-reversible transformation is solved similarly. The result is:
\emph{
$\psi\xrightarrow{G-cov}\phi$ iff there exists a  positive definite function $f(g)$ \cite{Pos-Def} such that $\chi_{\psi}(g)=\chi_{\phi}(g)f(g)$ for all $g\in G$.}

Finally, it is interesting to consider the asymmetry properties of $N$ copies of state $\psi$, for arbitrarily large $N$.   Again, we identify the asymmetry properties by considering interconvertability of states.  One difference with the single-copy case, however, is that we allow the conversion to be approximate, as long as the error goes to zero in the limit of arbitrarily many copies.   We say that there exists an \textbf{asymptotic G-covariant} transformation from state  $\psi$ to $\phi$  at rate $R\left(  \psi\rightarrow\phi\right)$ iff  $\psi^{\otimes N}\xrightarrow{G-cov} \phi_{M(N)}$ such that $\lim_{N\rightarrow\infty}\textrm{Fid}\left(\phi_{M(N)},\phi^{\otimes M(N)}\right)=1$ where $M(N)=\lfloor N R \left(  \psi\rightarrow\phi\right)\rfloor$ and $\textrm{Fid}\left(\psi_{1},\psi_{2}\right)$ is the fidelity between $\psi_{1}$ and $\psi_{2}$  \cite{Fid}.
We say that there exists a \textbf{reversible asymptotic G-covariant} transformation from $\psi$ to $\phi$ at rate $R\left(  \psi\rightarrow\phi\right)$ if there is an asymptotic G-covariant transformation from $\psi$ to $\phi$ at rate $R=R\left(  \psi\rightarrow\phi\right)$ and an asymptotic transformation from $\phi$ to $\psi$ at rate $R^{-1}$. 

 As it turns out, to specify the asymmetry properties in this case one requires less information about the state than is contained in $\chi_{\psi}(g)$. 
 Let $\{L_{k}\}$ be a basis for the Lie algebra $\mathfrak{g}$ associated to the compact Lie group $G$. Then we define the covariance matrix of the state $\psi$ as
$[C_{\mathfrak{g}}]_{kl}\left(  \psi\right)  \equiv1/2\left\langle \psi|{L}_{k}{L}_{l}+
{L}_{l}{L}_{k}|\psi\right\rangle -\left\langle \psi|{L}_{k}\left\vert
\psi\right\rangle \left\langle \psi\right\vert {L}_{l}|\psi\right\rangle$.
Now we can state the result:
\emph{
For a compact Lie group  G, if there exists a reversible asymptotic G-covariant transformation between $\psi$ and $\phi$  at rate $R\left(  \psi\rightarrow\phi\right)$ then
i)  $\textrm{Sym}_G(\psi)=\textrm{Sym}_G(\phi)$,
ii) $C_{\mathfrak{g}}\left(  \psi\right)= R\left(  \psi\rightarrow\phi\right) C_{\mathfrak{g}}\left(  \phi\right)$, iii)  ${\langle L \rangle_\psi}=R\left(  \psi\rightarrow\phi\right)  {\langle L\rangle_\phi}$ for  $L$ any arbitrary element of the commutator subalgebra  $[\mathfrak{g},\mathfrak{g}]$ (See \cite{derived}).
}
We conjecture that (i)-(iii) are also sufficient if the group is connected.
What is the interpretation of these three conditions? Since the characteristic function of $\psi^{\otimes N}$ is $\chi^{N}_{\psi}(g)$ (by the multiplicative property of characteristic functions) and $\chi_{\psi}(g)\le 1$, then at the limit of large $N$ the characteristic function of $\psi^{\otimes N}$ is almost zero everywhere in $G$ except at the neighbourhood around the points of  $\text{Sym}_{G}(\psi)$.  So to specify the asymmetry properties of $\psi^{\otimes N}$ we need to know $\text{Sym}_{G}(\psi)$ and the first and the second derivatives of $\chi_{\psi}(g)$ at these points. The covariance matrix specifies the second derivatives and the expectation value of the generators specifies the first derivatives. 

It is worth mentioning that the covariance matrix $C_{\mathfrak{g}}\left(  \psi\right)$ is proportional to the Fisher information matrix for the set  $\{U(g)|\psi\rangle\}$ at point $g=e$ and so condition $(ii)$ can be interpreted from the information theoretic point of view as conservation of information in reversible transformations. On the other hand, condition $(iii)$ can be understood in the dynamical point of view as a generalization of conservation of angular momentum.

In conclusion, in this letter we have introduced a general framework for the study of asymmetry of states with respect to an arbitrary  finite or compact Lie group. In particular,  we have focused on the asymmetry properties of a pure state and shown that these can be specified by the characteristic function of the state over the group. We have found the necessary and sufficient conditions for transforming one pure state to another by deterministic reversible and deterministic irreversible G-covariant operations and necessary conditions for asymptotic interconversion to be possible at a given rate.  Also, we have introduced the idea of the duality between the dynamical and information-theoretic perspectives on the consequences of dynamical symmetries, which we expect to have many significant applications.


\textbf{Acknowledgements:}
We acknowledge  G. Chiribella, G. Gour and S. Croke for helpful discussions and comments on the manuscript. Perimeter Institute is supported by the Government of Canada through Industry Canada and by the Province of Ontario through the Ministry of Research and Innovation. I. M. is supported by a Mike and Ophelia Lazaridis fellowship.


\begin{thebibliography}{000}
\bibitem{Fid} Fidelity is a measure of distinguishability of two states defined as $\textrm{Fid}(\rho,\sigma)\equiv\textrm{tr}(\sqrt{\sqrt{\rho}\sigma\sqrt{\rho}})$.
\end{thebibliography}
\end{document}